\documentclass[11pt]{article}
\usepackage{amssymb}
\usepackage{wasysym}
\newcommand{\be}{\begin{eqnarray}}\newcommand{\beq}{\begin{equation}}
\newcommand{\ee}{\end{eqnarray}}\newcommand{\eeq}{\end{equation}}
\newcommand{\ep}{\epsilon}\newcommand{\eps}{\varepsilon}
\newcommand{\la}{\lambda}\newcommand{\De}{\Delta}
\usepackage{epsf}
\usepackage{graphicx} 
\title
{
Temperature effect on the small-to-large crossover lengthscale of 
hydrophobic hydration 
}
\author{Y. S. Djikaev\thanks{
Corresponding author. 
E-mail: idjikaev@buffalo.edu
}
\hspace{0.1cm} and \hspace{0.1cm}E. Ruckenstein
\hspace{0.2cm}  \\ 
\\ Department of Chemical and Biological  Engineering, SUNY at Buffalo, \\ 
Buffalo, New York  14260 }
\date{ \hfill }
\renewcommand{\baselinestretch}{2}
\begin{document}
\renewcommand{\baselinestretch}{1}
\maketitle
\renewcommand{\baselinestretch}{1}
\vspace{-0.6cm}   
{\bf Abstract.} 
{\small 

The thermodynamics of hydration is expected to change gradually from entropic for small solutes to enthalpic for
large ones. The small-to-large crossover lengthscale of hydrophobic hydration depends on the thermodynamic
conditions of the solvent such as  temperature, pressure, presence of additives, etc...  We attempt to shed some
light on the temperature dependence of the crossover lengthscale by using a probabilistic approach to water
hydrogen bonding that allows one to obtain an analytic expression for the  number of bonds per water molecule as
a function of both its distance to a solute and solute  radius.  Incorporating that approach into the density
functional theory, one can examine  the solute size effects on its hydration over the entire small-to-large
lengthscale range at a series of different temperatures.   Knowing the dependence of the hydration  free energy
on the temperature and solute size, one can also obtain its enthalpic and entropic contributions  as functions of
both temperature and solute size. These function can provide some interesting insight into the temperature
dependence of the crossover lengthscale of hydrophobic hydration. The model was applied to the hydration of
spherical particles of various radii  in water in the temperature range from $T=293.15$ K to $T=333.15$ K. The
model predictions for the temperature dependence of the hydration  free energy  of small hydrophobes are
consistent with the experimental and simulational data on the hydration of simple molecular solutes. Three
alternative definitions for the small-to-large crossover length-scale of hydrophobic hydration are proposed, and
their temperature dependence is obtained. Depending on the definition and temperature,  the small-to-large
crossover in the hydration mechanism is predicted to occur for hydrophobes of radii from one to several
nanometers. Independent of its definition,  the crossover length-scale is predicted to decrease with increasing
temperature. 

}

\newpage 
\section{Introduction}
\renewcommand{\baselinestretch}{2}

Hydrophobic effects, i.e., hydrophobic hydration and hydrophobic interactions, are believed to  constitute an
important (if not crucial) element of a wide variety of  physical, chemical,  and  biological
phenomena,$^{1-10}$   such as the immicibility of oil and water, micelle and membrane formation,  the formation,
stability, and unfolding of the  native structure of a biologically active protein, etc... Hydrophobic hydration
is the thermodynamically unfavorable dissolution of a hydrophobic particle  (microscopic or macroscopic),
whereof the accommodation in water is accompanied by an increase in the associated free energy due to 
structural (and possibly energetic) changes in water around the hydrophobe. Since  the total volume of water
affected by two  hydrophobes is smaller when they are close together than when they far away from each other,
there appears an effective, solvent-mediated attraction between them which is referred to as hydrophobic
attraction. Most properties of hydrophobic interactions may be unambiguously determined from the analogous
properties of hydrophobic hydration; the former can be regarded as a partial reversal of the latter.

Both hydrophobic hydration and hyrophobic interactions have been the subject of intensive theoretical,
simulational, and experimental research for several decades.$^{11-14}$  Still the understanding of many aspects 
of these
effects remains rather unsatisfactory (sometimes even contradictory visions  of the same issue arise from
different research reports). For example, although the dependence of hydrophobic  effects on both the solvent
temperature and the hydrophobe size is not contested,$^{}$ many thermodynamic and molecular details thereof are
still to be elucidated.$^{5-8,15-17}$ 

The temperature and length-scale dependence of hydrophobic effects clearly transpires in two  exciting problems
of modern biophysics, namely, protein folding and protein denaturation.$^{5-8,15-18}$ Upon folding, a protein
buries its nonpolar amino acids into a globular core, away from contact with water, to form a  ``native"
configuration in which the protein is biologically functional; the burying of amino acids occurs in a particular
temporal and spatial order depending, among other factors, on their sizes.  The strength of hydrophobic
interactions largely determines the temperature range where the  native configuration remains stable (hence
protein physiologically active) before its thermal unfolding (denaturation) occurs. The (amino acid)
size-dependent  weakening of hydrophobic interactions at lower temperatures was suggested$^{8,18,19}$ to be an
important factor in the cold denaturation of proteins.

In simplified terms, the hydrophobicity of a solute particle can be regarded as a consequence of an unfavorable
entropy change that either overweighs a favorable energy change or supplements the unfavorable energy change, 
both occurring upon the accommodation of the particle in water. That is, the entropic contribution to the free
energy change upon hydration is always positive while the enthalpic contribution thereto can be either negative
(for small enough solutes) or positive (for larger ones).  The actual mechanism  of hydrophobic 
effects depends on the size and nature of solute particles involved as well as on thermodynamic conditions
imposed on the solvent (temperature, pressure, etc... ).

The hydration of small hydrophobic molecules (of sizes comparable to a water molecule) is believed to be
entropically ``driven" (and so is their solvent-mediated interaction) at all temperatures.$^{13,14}$  Such
molecules can  fit into the water hydrogen-bond network  without destroying any bonds. While this  results in a
negligible enthalpy of  hydration, the solute  constrains some degrees of freedom of neighboring water molecules
which    gives rise to negative hydration entropy and hence to positive hydration free energy. However, such a
simple mechanism has recently come under scrutiny$^{13,14}$  because  there are simulations$^{20,21}$ and
theory$^{22}$ suggesting that, under some conditions, the hydration of small hydrophobic molecules could be
entropically favorable. 

The hydration of large hydrophobic particles is believed to occur via a different  mechanism.$^{13,14,23,24}$
When inserted into liquid water, a large hydrophobe breaks some hydrogen bonds in its vicinity.  This would
result in large positive hydration  enthalpy and hence in a  free energy change proportional to the solute
surface area (as opposed to being proportional to the solute volume for small hydrophobes). Thus, the hydration
of large hydrophobic particles is  expected to be enthalpically driven (and so is their solvent-mediated
interaction). 

As the thermodynamics of hydration is expected to change gradually from entropic for small solutes to enthalpic
for large solutes, so are the structural properties of liquid water in the vicinity of the 
solutes.$^{13,14,18,25-29}$ The small-to-large crossover lengthscale is expected to  depend on thermodynamic
conditions of the solvent (as well as on the nature of the hydrophobe),  such as temperature, pressure, presence
of additives, etc... Its dependence on external pressure and concentration of additives was investigated in
ref.28 where it was shown to be nanoscopic under  ambient conditions but to decrease to molecular sizes
upon applying hydrostatic tension or adding ethanol to the solvent (water). On the other hand, 
the temperature effects on the crossover lengthscale had until recently received less than due
attention.$^{30,31}$
In the present manuscript we further investigate this particular  aspect of multifaceted hydrophobic phenomena. 


In order to shed some light on the temperature dependence of the crossover lengthscale of hydrophobic
hydration, we will use our previously developed probabilistic hydrogen bond (PHB)  approach$^{32}$ to
water hydrogen bonding that allows one to obtain an analytic expression for the  number of bonds per water
molecule as a function of both its distance to a hydrophobe and hydrophobe  radius.  Incorporating that
approach into the density functional theory (DFT), we will  examine the particle size effects on the hydration
of particles over the entire small-to-large  lengthscale range at solvent (water) different temperatures. For a
hydrophobe of a given size,  knowing the temperature dependence of its hydration free energy allows one to
calculate the enthalpic and entropic contributions thereto as functions of temperature. Thus, one can obtain
the free energy of hydration and its enthalpic and entropic contributions as functions of both temperature and
hydrophobe size. The analysis of these function can provide some interesting insight into the temperature
dependence of the crossover lengthscale of hydrophobic hydration. 

\section{Free energy of hydration and its energetic/enthalpic and entropic contributions}

The free energy of hydration of a hydrophobic solute of radius $R$ (the sphericity of the solute being assumed)
at temperature $T$ can be determined as a difference between the values of the appropriate 
free energy of the system (liquid water) with and without a hydrophobe therein. The free energy of interest must
be appropriate for the thermodynamic conditions under which the hydration process takes place.  
 
For example, if hydration occurs at constant temperature $T$, volume $V$, and number of molecules $N$ in 
the solvent (i.e., in a canonical ensemble),  the hydration free energy is 
\beq \De F_{\mbox{\tiny }}=F-F_0,\eeq 
where $F$ and $F_0$ are the Helmholtz free energies of the system (liquid water) with and 
without a hydrophobe therein, respectively. 
Likewise, if hydration occurs at constant temperature, volume, and chemical potential $\mu$ of 
the solvent (i.e., in a grand canonical ensemble),  the hydration free energy is 
\beq \De \Omega_{\mbox{\tiny }}=\Omega-\Omega_0,\eeq 
where $\Omega$ and $\Omega_0$ are the grand thermodynamic potentials  
of the system with and without a hydrophobic particle therein, respectively.

It should be noted that in the thermodynamic limit (of $N\rightarrow \infty, V\rightarrow \infty,
N/V=const$) the hydration free energy can be expected to be independent of whether 
hydration is canonical or grand canonical or another kind. Indeed, according to the thermodynamic 
theorem about small perturbations,$^{33}$ the corrections to the 
internal energy $E$, Helmholtz free energy $F$, grand 
thermodynamic potential $\Omega$, enthalpy $W$, and Gibbs free energy $G$ are equal to each other if they are 
calculated at constant variables indicated by the corresponding subscripts in 
\beq 
(dE)_{S,V,N}=(dF)_{T,V,N}=(d\Omega)_{T,V,\mu}=(dW)_{S,P,N}=(dG)_{T,P,N}=\sum\limits_{i}\Lambda_id\lambda_i,
\eeq
where $S$ and $P$ denote the entropy and external pressure of the system, whereas 
$dE, dF, d\Omega, dW$, and $dG$ are small changes in the thermodynamic potentials 
occurring in a quasistatic process due to external perturbations. 
The latter are represented in eq.(3) by the rightmost part, 
$\sum\limits_{i}\Lambda_id\la_i$, where $\Lambda_i$ is 
an intensive (field) variable conjugate to the external constraint $\la_i$ imposed on the system; e.g., such a
constraint and its conjugate field variable are responsible for a forced accommodation of a hydrophobic
solute in water upon its hydration. We will use 
this theorem in calculating the temperature dependence of the hydration free energy (see Section 4 below).

Knowing the free energy of hydrophobic hydration, one can find $\Phi_{\mbox{\tiny S}}$ and  
$\Phi_{\mbox{\tiny E}}$, the entropic and energetic contributions to $\De F$, as  
\beq 
\Phi_{\mbox{\tiny S}}\equiv-T\De S_{\mbox{\tiny }}=T(\partial \De F_{\mbox{\tiny }}/\partial T)_{\mbox{\tiny
V,N}},\;\;\;\;\; \Phi_{\mbox{\tiny E}}\equiv \De E_{\mbox{\tiny }}=(\partial (\De F_{\mbox{\tiny }}/T)/\partial
(1/T))_{\mbox{\tiny V,N}}, 
\eeq  
respectively, such that  $\De F_{\mbox{\tiny }}=\Phi_{\mbox{\tiny E}}+\Phi_{\mbox{\tiny S}}$ 
(in eq.(4) the subscripts of the partial derivatives
indicate the thermodynamic variables held constant upon taking the derivatives). 
Clearly, for the decomposition of $\De F_{\mbox{\tiny }}$ into energetic and entropic
components it is necessary to know its temperature dependence. This dependence is quite complicated (if not
impossible) to obtain by pure analytical means, 
but one can (numerically or simulationally) calculate the hydration free energy by applying either eq.(1) or
eq.(2) for a series of $T$'s and $R$'s and then contstruct the functions $\De F=\De F(R,T)$ and 
$\De\Omega =\De \Omega(R,T)$ by using interpolation.  

\section{Implementation of the probabilistic hydrogen bond model in the density functional theory}

The free energy of hydration of a hydrophobic particle of radius $R$ in water at temperature 
$T$ can be determined by using either computer simulations or DFT.$^{34-36}$  
As an illustration of the probabilistic hydrogen bond model, let us outline its implementation into DFT which
allows one to {\em explicitly} take into account the effect of water-water hydrogen bonding on the
hydrophobe-fluid interactions.

\subsection{Weighted density approximation of the density functional theory}

Assume that hydration takes place in an open system of constant  $\mu$, $V$, and $T$ 
(grand canonical ensemble). The hydration free energy is then found by using eq.(2). 
In DFT, the grand thermodynamic potential $\Omega$ of a nonuniform single component 
fluid, subjected to an external potential  $U_{\mbox{\tiny ext}}$ (representing a spherical 
hydrophobe of radius $R$), is a functional of the number density $\rho(\bf{r})$ of fluid molecules 
\be \Omega[\rho(\bf{r})]&=&\mathcal{F}_{\mbox{\tiny h}}[\rho({\bf r})] 
+ \frac1{2}\int\int d{\bf r} d{\bf r'}\,\rho({\bf r})\rho({\bf r'}) 
\phi_{\mbox{\tiny at}}(|\bf{r}-\bf{r'}|)\nonumber \\
&+&\int d {\bf r}\, U_{\mbox{\tiny ext}}({R,\bf r})\rho({\bf r})-
\mu\int d{\bf r}\, \rho({\bf r}),\ee
where $\mathcal{F}_{\mbox{\tiny h}}[\rho({\bf r})]$ is the intrinsic Helmholtz free energy
functional of hard sphere fluid, $\mu$ is the
chemical potential, and  $\phi_{\mbox{\tiny at}}(|\bf{r}-\bf{r'}|)$ is the attractive
part of the interaction potential between two fluid molecules located at ${\bf r}$ and ${\bf r'}$;
$U_{\mbox{\tiny ext}}({R,\bf r})$ is a 
the external potential whereto a fluid molecule is subjected near 
the hydrophobe; 
the integrals are taken over the volume $V$ of the system. Among various models for 
$\mathcal{F}_{\mbox{\tiny h}}[\rho({\bf r})]$, the weighted density approximation (WDA)$^{34,37-39}$ 
with a weight function independent of weighted density represents  
an optimal combination of
accuracy and simplicity. It is non-local with
respect to $\rho({\bf r})$; it takes into account short-ranged correlations and  captures 
the fluid density oscillations near a hard wall. 

The key element of WDA  is the weighted density $\widetilde{\rho}({\bf r})$ 
determined as a functional of $\rho({\bf r})$ via an implicit equation.$^{34,37,38}$ 
The  equilibrium density profile is obtained
by minimizing  $\Omega[\rho(\bf{r})]$ with respect to $\rho({\bf r})$.  The
corresponding Euler-Lagrange equation can be written as 
\be \mu=k_BT\ln(\Lambda^3\rho({\bf r}))+W({\bf r};\rho({\bf r})),\ee 
where $\Lambda=(h^2/2\pi mk_BT)^{1/2}$ is the thermal de Broglie wavelength of a molecule  of mass
$m$ ($h$ and $k_B$ being Planck's and Boltzmann's constants)  and 
$W({\bf r};\rho({\bf r}))$ is a function of ${\bf r}$ and a functional of $\rho({\bf r})$ 
(for more details see ref.39).$^{}$

The hydrophobe being spherical, the external potential is a function of a
single variable $x=r-R$, and the equilibrium density profile obtained from eq.(6) is a function of
a single variable $r$: $\rho({\bf r})=\rho(r)$. The substitution of $\rho(r)$ into eq.(5)
provides the grand thermodynamic potential $\Omega$ of the non-uniform fluid with a hydrophobe 
therein.  The grand thermodynamic potential  of uniform liquid water $\Omega_0$ 
without a hydrophobe therein can be found by setting $\rho({\bf r})=\rho_0$, where $\rho_0$ is the equilibrium
density of a uniform liquid at given thermodynamic conditions.

\subsection{Hydrogen bond contribution to the external potential $U_{\mbox{\tiny ext}}$} 

To apply either DFT or computer simulations to the thermodynamics of hydrophobic phenomena, it is
necessary to know the external potential field  $U_{\mbox{\tiny ext}}^{\mbox{\tiny }}\equiv
U_{\mbox{\tiny ext}}^{\mbox{\tiny }}(R,x)$. 
Usually, the interaction of fluid molecules with a foreign (impenetrable) substrate 
is treated in  the mean-field  approximation whereby  every fluid molecule is considered to be subjected to an
external potential, due to its pairwise interactions with the substrate molecules.$^{33,34}$  The
substrate effect  on the ability of fluid (water) molecules to form hydrogen bonds had been
previously  neglected.  However, using the PHB model,  one can explicitly implement that
effect in the DFT formalism. 

Indeed, the total external potential can be written as 
\beq U_{\mbox{\tiny ext}}=U_{\mbox{\tiny ext}}^{\mbox{\tiny p}}+
U_{\mbox{\tiny ext}}^{\mbox{\tiny h}},\eeq
where  $U_{\mbox{\tiny ext}}^{\mbox{\tiny p}}\equiv U_{\mbox{\tiny ext}}^{\mbox{\tiny p}}(R, x)$  represents the
(conventional) {\em pairwise} potential exerted by all the molecules, constituting the hydrophobe, on a water
molecule, and $U_{\mbox{\tiny  ext}}^{\mbox{\tiny h}}\equiv U_{\mbox{\tiny ext}}^{\mbox{\tiny h}}(R, x)$ is the 
water-water hydrogen bond contribution to $U_{\mbox{\tiny ext}}$.  
This contribution depends on the energy of a single hydrogen bond 
and number of bonds that a water molecule can form in the hydrophobe vicinity and in the bulk.$^{39,40}$ 
It can be determined as 
\beq U_{\mbox{\tiny ext}}^{\mbox{\tiny h}}=\frac1{2}(\eps_sn_s-\eps_bn_b).\eeq 
$n_b$ and $\eps_b$ are the number of hydrogen bonds per water molecule and the energy of a bond in the bulk, 
whereas $n_s$ and $\eps_s$ are the analagous quantitities in the hydrophobe vicinity. Although 
$U_{\mbox{\mbox{\tiny ext}}}^{\mbox{\tiny h}}$,  is due to  the 
deviation of $n_s$ from $n_b$  as well as  the deviation of $\eps_s$ from $\eps_b$; as previously,$^{39}$ 
the latter effect 
is neglected hereafter due to its uncertainty.$^{}$

The first term on the RHS of eq.(8) represents the total energy of hydrogen bonds  of a water molecule
at a distance $x$ from  the surface of a particle of radius $R$,  whereas the second term is the
energy of its hydrogen bonds in bulk (at $x\rightarrow \infty$); the factor $1/2$ is needed to
prevent  double counting the energy because every   hydrogen bond and its energy, either $\eps_s$ or
$\eps_b$, are shared between two  molecules. Note that the $R$ and $x$ dependence of 
$U_{\mbox{\mbox{\tiny ext}}}^{\mbox{\tiny h}}$ is determined by
the $R$ and $x$ dependence of $n_s$, while the temperature dependence of 
$U_{\mbox{\mbox{\tiny ext}}}^{\mbox{\tiny h}}$ is determined by the temperature dependence of 
$n_b$ and $\eps_b$ (see section 4).

\subsection{Number of hydrogen bonds per water molecule near a spherical hydrophobic surface} 

Consider a spherical hydrophobic  particle of radius $R$ immersed in liquid water.  Even if one assumes that the
intrinsic hydrogen bonding ability of a water molecule is not affected by the  hydrophobe, in its vicinity  a
``boundary" water molecule forms a smaller number  of bonds than in bulk  because the surface restricts the
configurational space  available to other water molecules  necessary for a boundary water molecule to form
hydrogen bonds. The  probabilistic model allows one to obtain an analytic expression for the average number of
bonds that a BWM can  form as a function of its distance to the hydrophobe and hydrophobe radius. 

In the probabilistic hydrogen bond model,$^{41}$  a water molecule is considered to have four arms each capable
of forming a single hydrogen bond. The configuration of four hydrogen-bonding (hb) arms is rigid and symmetric
(tetrahedral) with the inter-arm angles  $\alpha=109.47^\circ$.  Each hb-arm can  adopt a continuum of
orientations  subject to the constraint of tetrahedral rigidity. A water molecule can form a hydrogen bond with 
another molecule only when the tip of any of its hb-arms coincides with  the center of the  second molecule. The
length of a hb-arm thus equals the  length of a hydrogen bond $\eta$, assumed independent of whether the
molecules are in bulk or near a  hydrophobe.   The characteristic  length of pairwise interactions between water
molecules and  molecules constituting the hydrophobe is also assumed to be $\eta$.  

The location of a water molecule is determined by the distance $r$ from its center to  the center of the
hydrophobe which is also chosen as the origin of the spherical coordinate system. The distance $x$ between water
molecule and hydrophobe is defined as $x=r-R$. A spherical layer of thickness $\eta$ from $r=R+\eta$ to 
$r=R+2\eta$ is referred to as the solute hydration layer (SHL). 

The number of hydrogen bonds per water molecule near the hydrophobe depends on both $R$ and $x$, 
i.e., $n_s=n_s(R,x)$ (the dependence of $n_s$ on $T$ is discussed in section 4). the function $n_s(R,x)$  attains its minimum at $x=\eta$,  because at this
distance the configurational space available for neighboring water molecules is most restricted  compared to bulk
water. On the other hand,  if $x>2\eta$, the number of
hydrogen bonds that the water  molecule can form is assumed to be unaffected by the hydrophobe:  $n_s(R,x)=n_b$
for $x\ge 2\eta$.  Thus,  according to eq.(8), 
$U_{\mbox{\tiny ext}}^{\mbox{\tiny h}}(R,x)\ne 0$ only for  $\eta\le x \le 2\eta$. 

In the framework of the PHB approach,$^{41}$ the function $n_s=n_s(R,x)$ is represented as  
\beq n_s=k_1b_1+k_2b_1^2+k_3b_1^3+k_4b_1^4,\eeq 
where $b_1$ is the probability that one of the hb-arms (of a bulk water molecule) can form a 
hydrogen bond and the coefficients  $k_1,k_2,k_3$, and $k_4$ depend on $R$ and $x$, and so does 
$n_s$. Equation (9) assumes that  the {\em intrinsic} hydrogen-bonding 
ability of a BWM (the tetrahedral configuration of its hb-arms and their lengths and
energies) is unaffected by the hydrophobe, but it takes into
account  the  constraint that near the hydrophobe some  orientations of  the  hb-arms of a water molecule 
cannot lead to the formation of hydrogen bonds. This constraint depends on the 
distance between water molecule and hydrophobe and on the hydrophobe radius, whence the $R$- and $x$-dependence 
of $k_1, k_{2}, k_{3},$ and $k_{4}$.

The functions $k_1\equiv k_1(R,x),k_2\equiv k_2(R,x),k_3\equiv k_3(R,x)$, and $k_4\equiv k_4(R,x)$
can be  evaluated by using geometric considerations.$^{30}$  They all become equal to $1$ at $x\ge
2\eta$, where eq.(9) reduces to its bulk analog,  $n_b=b_1+b_1^2+b_1^3+ b_1^4$ (see ref.39). Since 
experimental data on $n_b$ are readily available, one can
find $b_1$ as a positive solution (satisfying $0<b_1<1$)  of the latter equation. 
Thus, equation (9) provides an efficient pathway to $n_s$ as a function of $x$ and $R$ (as well as $T$, see the
next Section).


\section{Numerical Calculations}

For a numerical illustration, we considered the hydration of spherical hydrophobes of radii 
$R/\eta=1,3,5,7,10,15,20,30,50,100$ in the model water at five temperautres, $T=293.15$ K, $303.15$ K,  $313.15$
K, $323.15$ K,and $333.15$ K, the chemical potential being the same at all temperatures, $\mu=-11.5989$ $k_BT_0$
(with $T_0=293.15$ K). Hydration was assumed to occur at constant $\mu, V, T$, so that  
the hydration free energy was determined as the change in the grand canonical
potential of the system, $\De \Omega$,  by using the combined PHB/DFT formalism as outlined above. 
According to the thermodynamic theorem about small corrections, eq.(3), one can then set $\De F\approx \De
\Omega$ and carry out the decomposition of the hydration free energy into its enthalpic and entropic
contributions by using eq.(4).

As clear from the foregoing, the temperature dependence of  $\De \Omega_{\mbox{\tiny }}$ (or $\De F$) contains a
contribution from the temperature dependence of   $U_{\mbox{\tiny ext}}^{\mbox{\tiny h}}$, hydrogen bond
contribution to the total external field.  The dependence of $U_{\mbox{\tiny ext}}^{\mbox{\tiny h}}$ on $T$ is 
due to the temperature dependence of four quantities: $n_s,\,n_b,\, \ep_s$, and $\ep_b$ (see eq.(8)). The
functions  $\ep_b\equiv \ep_b(T)$ and $n_b\equiv n_b(T)$ are either readily available or can be constructed on
the basis of available data.$^{}$ Therefore, by virtue of eq.(9), the dependence of $n_s$ on $T$  can be considered to be known as
well.  The energy of a hydrogen bond was assumed to depend on temperature in such a way
that $\ep_s(T)/\ep_b(T)\approx const$ in the temperature range considered.  One can thus consider $
U_{\mbox{\tiny ext}}^{\mbox{\tiny h}}$  to be a known function of not only $x$ and $R$, but also $T$: 
$U_{\mbox{\tiny ext}}^{\mbox{\tiny h}}=U_{\mbox{\tiny ext}}^{\mbox{\tiny h}}(R,x,T)$. This allows one to
numerically determine the temperature dependence of the free energy of hydration and to subsequently use
interpolation procedure to find an analytical fit thereof which then can be used in eq.(4). 

The liquid state of bulk water  was ensured by imposing the appropriate boundary condition onto eq.(6),
$\rho(x)\rightarrow \rho_l$ as $x\rightarrow \infty$, with $\rho_l$ the bulk liquid density.  The densities 
$\rho_v$ and $\rho_l$ of coexisting vapor and liquid, respectively, are determined  by solving the equations 
$\left.\mu(\rho,T)\right|_{\rho=\rho_v}=\left.\mu(\rho,T)\right|_{\rho=\rho_l},\;\;\;\;\;\;\;\;
\left.p(\rho,T)\right|_{\rho=\rho_v}=\left.p(\rho,T)\right|_{\rho=\rho_l}$, requiring  the chemical potential
$\mu\equiv\mu(\rho,T)$ and pressure  $p\equiv  p(\rho,T)$ to be the same throughout both coexisting phases. The
liquid densities for the above  five temperatures thus obtained were
$\rho\eta^3=0.6342,\,0.6517,\,0.6647,\,0.6750$, and $0.6835$,  respectively. 

Solving eq.(6), the chemical potential of a uniform hard sphere fluid  $\mu_{\mbox{\tiny h}}$  and the 
configurational part  $\Delta\psi_{\mbox{\tiny h}}\equiv\Delta\psi_{\mbox{\tiny h}}(\rho,T)$ of the free energy
of a hard sphere fluid were modeled in  the Carnahan-Starling approximation,$^{35,36,42}$  whereas for the weight
function $w(|{\bf r'}-{\bf r}|;\widetilde{\rho}({\bf r}))$ (entering in the implicit equation that determines
$\widetilde{\rho}({\bf r}))$ as a functional of $\rho({\bf r})$)  we adopted a  $\widetilde{\rho}$-independent
version,$^{38}$  
$$\mu_{\mbox{\tiny h}}=k_BT\Big(\ln(\Lambda^3\rho)+\xi\frac{8-9\xi+3\xi^2}{(1-\xi)^3}\Big),\;\;\;
\Delta \psi_{\mbox{\tiny h}}=k_BT\frac{\xi\,(4-3\xi)}{(1-\xi)^2},\;\;\;\;\; 
w(r_{12})=\frac3{\pi\eta^4}(\eta-r_{12})\Theta(\eta-r_{12}),$$   
with $\xi=(\pi d^3/6)\rho$ and $\Theta(u)$ being the Heaviside (unit-step) function (the quantities 
$\mu_{\mbox{\tiny h}}$, $\Delta\psi_{\mbox{\tiny h}}$, and $w(|{\bf r'}-{\bf r}|;\widetilde{\rho}({\bf r}))$ are 
all needed$^{32,37-39}$ to calculate $W({\bf r};\rho({\bf r}))$ in eq.(6)).

The pairwise  interactions of water molecules were modeled by using the Lennard-Jones (LJ) potential with the
energy parameter  $\eps_{\mbox{\tiny ww}}=3.79\times 10^{-14}$ erg  and the diameter $d$ of a model  molecule 
set to be $\eta$.  The attractive part $\phi_{\mbox{\tiny at}}$ of pairwise water-water interactions was modeled
via the Weeks-Chandler-Anderson perturbation scheme.$^{43}$  The interaction potential between water molecule and
molecule of a  hydrophobe was assumed to be of LJ type with an energy parameter $\eps_{\mbox{\tiny wp}}$ and a
length parameter $\eta$.  Integrating this interaction with respect to the position of the molecule of the
hydrophobe over the hydrophobe volume $V_R=4\pi R^3/3$, one can obtain  the pairwise contribution $U_{\mbox{\tiny
ext}}^{\mbox{\tiny p}}$ into $U_{\mbox{\tiny ext}}$.  The dimensionless number density of molecules in
the hydrophobe was set to be $\rho_p\eta^3\approx 1$.  The 
density profiles and hydration free energies and its enthalpic and entropic component thus obtained are shown in
Figures 1 through 5.

Figure 1 presents the density profiles near a spherical hydrophobe of 
radius $R$ with the degree of hydrophobicity $\eps_{\mbox{\tiny
wp}}/\eps_{\mbox{\tiny ww}}=0.75$. Different figure panels show results for
different radii ($R/\eta=1,3,5,7,10,15,30,100$). In each panel, different curves correspond to different temperatures
($T=293.15, 303.15, 313.15, 323.15, 333.15$ from bottom to top, respectively). 
As clear, both the temperature and the hydrophobe radius greatly affect 
the distribution of vicinal water molecules. 

At lower temperatures 
$T=293.15 K$, and $303.15$ K, the oscillations in the density profile gradually disappear as  $R$ 
increases. They  are well pronounced for $R/\eta=1$, but virtually 
non-existent for particles $R/\eta\ge 7$.  As $R$ increases, a thin
depletion layer around the particle  (virtually non-existent for $R/\eta=1$) becomes more 
developed, with its density approaching that of vapor and its thickness approaching $\eta$. 
However, at higher temperatures, the density oscillations are present in the vicinity of hydrophobes of all
sizes. On the other hand, the higher the temperature the narrower the thickness of the depletion layer near a 
hydrophobe and the higher the fluid density therein. These latter effects have a clear lengthscale dependence.

Indeed, for a hydrophobe of $R/\eta=1$ the temperature increase from $T=293.15$ to $T\gtrsim 303.15$ leads to
the increase of the ``contact" fluid density (i.e., the fluid density at contact with the hydrophobe)  by many
orders of magnitude, i.e., the vapor-like depletion layer transforms into a liquid like depletion layer whereof
the thickness also quickly decreases with increasing temperature. At $T=333.15$ K the contact fluid density
equals about $0.5\rho_l$  and the width of the depletion layer equals about $0.3\eta$; the corresponding
quantities at $T=293.15$ K are equal to about $0.01\rho_l$ and $0.7\eta$. 
On the other hand, for a hydrophobe of
$R/\eta=3$, at $T=333.15$ K  the contact fluid density equals about $0.02\rho_l$  and the width of the depletion
layer equals about  $0.67\eta$, whereas at $293.15$ the corresponding quantities are about $0.0003\rho_l$ and
$0.88\eta$. Thus, the temperature dependence of these two characteristics of the fluid density profile near a
hydrophobe is very sensitive to the hydrophobe size. This sensitivity is consistent with the largely accepted
view that the underlying physics of hydrophobicity is different on different length scales. 
It should be noted that characterizing the hydrophobicity of a solute by means of 
the 
water density profile in its vicinity is freight with ambiguities.$^{44}$ 
It was shown that the water density near weakly hydrophobic surfaces can
be bulk-like and hence cannot serve as a foundation to quantify the surface hydrophobicity; instead, 
liquid water density fluctuations (or the probability of cavity formation) would constitute  a more
adequate quantitative measure of the surface hydrophobicity.$^{44}$ Our model in its present  form
does  not contain provisions to capture the fluctuations of the number of hydrogen bonds per water molecule, but
its appropriate modification is possible and one can expect that such fluctuations, besides depending on $R$ and
$x$, will be strongly correlated with the hydrophobicity of the solute particle, i.e., will depend on the ratio
$\eps_{\mbox{\tiny wp}}/\eps_{\mbox{\tiny ww}}$ as well.

Figure 2a, 2b, and 2c present the free energy of hydrophobic  hydration and its entropic and
enthalpic constituents, respectively,  vs temperature (note that 
the curves are provided only for guiding the eye; the actual
calculated points are at $T=293.15, 303.15, 313.15, 323.15, 333.15$).
The curves correspond to 
$R/\eta=1,3,5,7,10,15,20,30,50$, and $100$ (from bottom to top in Figs.2a and 2c, from top to bottom in Fig.2b). 
The intrinsic hydrophobicity of the
particles is assumed to be independent  of $R$, with  $\eps_{\mbox{\tiny wp}}/\eps_{\mbox{\tiny ww}}=0.75$.
The hydration free energy and its constituents are expressed in  units of $k_BT_1$ per 
``dimensionless unit area"; the dimensionless quantities 
$\overline{\De\Omega}, \overline{\Phi_S}$, and $\overline{\Phi_E}$ are defined as 
$\overline{\De\Omega}=\De \Omega/(k_BT_1(4\pi R^2/\eta^2))$, 
$\overline{\Phi_S}=\Phi_S/(k_BT_1(4\pi R^2/\eta^2))$,and 
$\overline{\Phi_E}=\Phi_E/(k_BT_1(4\pi R^2/\eta^2))$,
respectively. 

In the considered temperature range from 293.15 K to 333.15 K both the hydration
free energy and its entropic constituent increase with temperature for small hydrophobes (with $R/\eta=1,3,5$) 
but decrease for larger ones (with $R/\eta=7,10,15,20,30,50,100$; the energetic contribution to the free energy
is virtually independent of temperature for very small and very large hydrophobes ($R/\eta=1,3,20,30,50,100$) but and an increasing function of temperature for all hydrophobes . 
increases with the temperature in the small-to-large range of hydrophobe length scales ($R=5,7,10,15$).

Figures 3a, 3b, and 3c present the free energy of hydrophobic  hydration  $\De F_{\mbox{\tiny }}$  and its
enthalpic and entropic constituents, respectively, as  functions of the hydrophobe radius $R$ (note that the
curves are provided only for guiding the eye; the actual calculated points are at
$R/\eta=1,3,5,7,10,15,20,30,50$, and $100$).  The intrinsic hydrophobicity of the particles is assumed to be
independent  of $R$, with  $\eps_{\mbox{\tiny wp}}/\eps_{\mbox{\tiny ww}}=0.75$. The hydration free energy and
its constituents are shown as the dimensionless quantities  $\overline{\De\Omega}, \overline{\Phi_S}$, and
$\overline{\Phi_E}$.  There are five different curves in Figs.3a-3c corresponding to five different temperatures, 
$T=293.15, 303.15, 313.15, 323.15, 333.15$ (from bottom to top in Figs.3a and 3b and from top to). 

The variable  sensitivity of $\overline{\De\Omega}, \overline{\Phi_S}$, and $\overline{\Phi_E}$  to $R$ is
another indication that the hydration of small and large length-scale particles occur via different  mechanisms.
For small hydrophobes,  they vary sharply with increasing $R$, but become weakly sensitive to $R$ for large
enough hydrophobes; for hydropobes with $R/\eta \gtrsim 15$ 
the curves for different temperatures are hardly discernable from each other (particularly in the case of 
$\overline{\Phi_E}$ which is virtually temeperature independent for $R/\eta \lesssim 5$).

As expected, the always-positive hydration free energy $\overline{\De\Omega}$ increases
with increasing $R$, quite sharply for small hydrophobes (with $R/\eta \lesssim 5$), but relatively weakly for
large ones  (with $R/\eta \gtrsim 15$). Its entropic contribution $\overline{\Phi_S}$ behaves
in the inverse (decreasing with increasing $R$) manner on the same length scales; 
it is a decreasing function of $R$, being largely positive for small hydrophobes but sharply
decreasing with increasing $R$, crossing zero at some $R_S^{0}$, and then becoming negative and a week 
function of $R$. 
The enthalpic  contribution
$\overline{\Phi_S}$ is an increasing function of $R$. For small hydrophobes  it is negative but sharply
increases with increasing $R$, crossing zero at some $R_E^{0}$, and then becoming positive and a week 
function of $R$.   
Note that in the region of negative
$\overline{\Phi_E}$, i.e. for $0<R\lesssim R_E^{0}$, both contributions are close by absolute value with the
entropic contribution just slightly dominant, which leads to relatively small value of the hydration free energy
even there where its components $\overline{\Phi_E}$ and $\overline{\Phi_S}$  have large (negative and positive,
respectively) values. This represents a well-established characteristic feature of the hydration of molecular
scale apolar solutes.$^{1,12}$  

Furthermore, there exists such a radius $R_{SE}$ that $\overline{\Phi_E}\lesseqgtr
\overline{\Phi_E}$ for   $R\lesseqgtr R_{SE}$. 
For hydrophobes of radii smaller than $R_{SE}$ the hydration free energy is dominated
by its entropic component ($\overline{\Phi_E} < \overline{\Phi_E}$), whereas for hydrophobes of radii larger
than $R_{SE}$ the hydration free energy is dominated by its enthalpic constituent ($\overline{\Phi_E} > 
\overline{\Phi_S}$). The radius $R_{SE}$ can be determined as a solution of the equation 
$\overline{\Phi_E}(T,R)=\overline{\Phi_S}(T,R)$.  

Clearly, the three ``crossover" radii $R_S^{0}$, $R_E^{0}$ and $R_{SE}$ are functions of temperarture:  
$R_S^{0}=R_S^{0}(T),\;R_S^{0}=R_S^{0}(T),\;R_{SE}=R_{SE}(T)$. Figure 4 shows these functions 
for the hydration of solutes of given 
hydrophobicity $\eps_{\mbox{\tiny wp}}/\eps_{\mbox{\tiny ww}}=0.75$ in the temperature range from $293.15$ K to
$333.15$ K (the curves are provided only for guiding the eye; the actual
calculated points are at $T=293.15, 303.15, 313.15, 323.15, 333.15$).  
Note that the $T$-dependence of $R_S^{0}$, $R_E^{0}$, and $R_{SE}$ is based on an 
approximation of the $T$-dependence 
of the hydration free energy; a more accurate approximation for the latter will certainly affect the functions  
$R_S^{0}=R_S^{0}(T),\;R_S^{0}=R_S^{0}(T),R_{SE}=R_{SE}(T)$ as well. 
Taking into account that $\eta\approx 3\times 10^{-8}$ cm, one can conclude that 
the predictions of our model are qualitatively consistent with the
recently reported estimate$^{30,31}$ for the crossover length scale; defining the latter as the solute size at
which the hydration entropy crosses from negative to positive (or vice versa) and analyzing the experimental
results obtained by Li and Walker$^{30}$ via single-molecule force spectroscopy, 
Garde and Patel$^{31}$ evaluated the crossover radius $R_S^{0}$ to be of the order of 1 nm at room 
temperatures. 

The temperature dependence of the hydration free energy predicted by the combined PHB/DFT model for small scale
hydrophobes  is in a reasonable agreement with both experimental data and simulational results reported for
simple solutes.$^{45-47}$ Figure 5 presents such a comparison for the temperature dependence of the  hydration
free energy. The three short solid lines correspond to the predictions of the PHB/DFT model  for solutes of radii
$R/\eta=1,\,3,\,5$ (from bottom to top, respectively) and of hydrophobicity  $\eps_{\mbox{\tiny
wp}}/\eps_{\mbox{\tiny ww}}=0.75$. The filled circles and squares represent the experimental data$^{45}$ for the
molecules of xenon and methane, respectively. The results obtained$^{46}$   by molecular  dynamics simulations
for the hydration of neon, argon, krypton, xenon, and methane molecules in an SPCE model water are shown as
short-dashed, long-dashed, short-long-dashed, dash-dotted, and dotted curves, respectively. Converting the
original data from refs.45 and 46 to the dimensionless hydration free energy per dimensionless  unit area, we set
the radii of the solute molecules to their corresponding van der Waals radii as provided on the web site
www.wikipedia.org; the ratio $R/\eta$ thus was $0.46$ for neon, $0.59$ for argon, $0.637$  for krypton, $0.68$
for  xenon, and $0.644$ for methane. Note also that the discrepancies between the simulated data and the PHB/DFT 
predictions can be partially accounted for by the differences in the ratio   $\eps_{\mbox{\tiny
wp}}/\eps_{\mbox{\tiny ww}}=0.75$ used in the latter and the corresponding ratios  used in the simulations
($0.237$ for neon, 1.598 for argon, 2.16 for krypton, 2.75 for xenon, and 1.89 for methane). Taking into account
these caveats as well as the discrepancies between the results obtained via MD simulations for different water
models$^{43}$ (SPC, SPCE, TIP3P, TIP4P, TIP5P), one can deem the predictions of the PHB/DFT model in
qualitatively satisfactory agreement with both experimental and simulational analogues.

\section{Conclusions}

The thermodynamics of hydration is expected to change gradually from entropic for small hydrophobic  solutes to
enthalpic for  large ones. The range of solute linear sizes over which the crossover of the hydration regime  is
expected to occur depends not only on the nature of the hydrophobe but also  on the thermodynamic conditions of 
the solvent such as  temperature, pressure, presence of additives, etc... 

In order to elucidate some molecular and thermodynamic aspects of the crossover properties of hydrophobic
hydration, including its crossover lengthscale, we have  used the combination of our previously developed
probabilistic hydrogen bond model with the density functional theory. The probabilistic approach  to water
hydrogen bonding allows one to obtain an analytic expression for the  number of bonds per water molecule as a
function of both its distance to a solute and solute  radius. This function  allows one to explicitly
incorporate the effect of water-water hydrogen bonding, crucial element of hydrophobic phenomena,  on the
water-hydrophobe interactions into the density functional theory.  One can thus examine the solute size effects
on its hydration over the entire small-to-large lengthscale range at a series of different temperatures. For a 
hydrophobe of given size,  knowing the temperature dependence of its hydration free energy allows one to
calculate the enthalpic and entropic contributions thereto as functions of temperature. Thus, one can obtain the
free energy of hydration and its enthalpic and entropic contributions as functions of both temperature and
hydrophobe size. The analysis of these function can provide some interesting insight into the temperature
dependence of the crossover lengthscale of hydrophobic hydration. 

As a numerical illustration of the combined PHB/DFT approach, we have studied the hydration of spherical
particles of various radii and of fixed hydrophobicity in a model water in the temperature range from 293.15 K to
333.15 K. The predictions for the temperature of the hydration  free energy of small size hydrophobes are
qualitatively and even quantitatively consistent with the experimental and simulational data on  the hydration of
simple molecular solutes (neon, argon, krypton, xenon, and methane).   We have shown that one can give three 
alternative definitions for the small-to-large crossover length-scale of hydrophobic hydration, based on the size
dependence of either hydration enthalpy or hydration entropy or enthalpy-entropy comparison.  According to all
three definitions, the crossover   in the hydration mechanism is predicted to occur for hydrophobes of radii from
one to several nanometers. The numerical  calculations have denonstrated that, although the  temperature
dependence of the crossover length-scale depends on its definition, all three definitions result  in the
crossover length-scale decreasing with temperature. The three crossover length-scales differ from each other most
at low temperatures and monotonically approach each other with increasing temperature. 

\section*{References}
\begin{list}{}{\labelwidth 0cm \itemindent-\leftmargin} 

\item $^{1}$K.A.Sharp,   Curr. Opin. Struct. Biol. {\bf1}, 171-174 (1991). 
\item $^{2}$W.Blokzijl and J.B.F.N.Engberts,   Angew. Chem. Int. Ed. Engl. {\bf32}, 1545-1579 (1993).
\item $^{3}$K.Soda,   Adv. Biophys. {\bf 29}, 1-54 (1993). 
\item $^{4}$M.E.Paulaitis and S.Garde, H.S.Ashbaugh,     
 Curr. Opin. Colloid Interface Sci. {\bf 1}, 376-383 (1996).
\item $^{5}$C.Ghelis and J.Yan,  {\it Protein Folding} (Academic Press, New York, 1982).
\item $^{6}$K.A.Dill,  Biochemistry {\bf 29}, 7133-7155 (1990). 
\item $^{7}$W.Kauzmann, Adv.Prot.Chem. {\bf 14}, 1-63 (1959).
\item $^{8}$P.L.Privalov, Crit.Rev.Biochem.Mol.Biol. {\bf 25}, 281-305 (1990).
\item $^{9}$A.Ben-Naim, {\it Hydrophobic interactions} (Plenum, New York, 1980).
\item $^{10}$C.Tanford, {\it The hydrophobic effect: Formation of micelles and biological membranes}  
(Wiley, New York, 1980). 
\item $^{11}$H.S.Ashbaugh, T.M.Truskett, and P.Debenedetti,  
 Phys.Chem.Chem.Phys. {\bf 116}, 2907-2921 (2002).
\item $^{12}$ a) B.Widom, P.Bhimulaparam, and K.Koga, Phys.Chem.Chem.Phys. {\bf 5},
	3085-3093 (2003);
	(b) K.Koga, P.Bhimulaparam, and B.Widom, Mol.Phys. {\bf 100}, 3795-3801 (2002).
\item $^{13}$P.Ball, P.  Chem.Rev. {\bf 108}, 74-108 (2008).
\item $^{14}$B.J.Berne, J.D.Weeks, and R.Zhou,  Annu.Rev.Phys.Chem. {\bf  60}, 85-103 (2009).
\item $^{15}$N.T.Southall, K.A.Dill, and A.D.J.Haymett,   J.Phys.Chem. B {\bf 106}, 521-33 (2002).
\item $^{16}$R.Zangi and B.J.Berne,  J.Phys.Chem.B {\bf 112}, 8634-8644 (2008).
\item $^{17}$D.Paschek,  J.Chem.Phys. {\bf 120}, 6674 (2004).
\item $^{18}$ a) D.M.Huang and D.Chandler, Proc.Natl.Acad.Sci. USA {\bf 97}, 8324 (2000);
(b) D.M.Huang and D. Chandler, J.Phys.Chem.B {\bf 106}, 2047-53 (2002). 
\item $^{19}$C.J.Tsai, J.V.Maizel, and R.Nussinov,  Crit.Rev.Biochem.Mol.Biol. {\bf 37}, 55 (2002).
\item $^{20}$K.Watanabe and H.C.Andersen,  J.Phys.Chem. {\bf 90}, 795-802 (1986).
\item $^{21}$C.Pangali, M.Rao, and B.J.Berne,  J.Chem.Phys. {\bf 71}, 2982-90 (1979).
\item $^{22}$L.R.Pratt and D.Chandler, J.Chem.Phys. {\bf 67}, 3683-3704 (1977).
\item $^{23}$F.H.Stillinger, J.Solut.Chem. {\bf 2}, 141-58 (1973).
\item $^{24}$C.Y.Lee, J.A.McCammon, and P.J.Rossky, J. Chem. Phys. {\bf 80}, 4448-55 (1984).
\item $^{25}$K.Lum, D.Chandler, and J.D.Weeks,  J.Phys.Chem.B {\bf 103}, 4570 (1999).
\item $^{26}$N.T.Southall and K.A.Dill, J.Phys.Chem. B {\bf 104}, 1326-1331 (2000).
\item $^{27}$L.R.Pratt, Annu. Rev. Phys. Chem. (2002) {\bf 53}, 409-36
\item $^{28}$S.Rajamani, T.M.Truskett, and S.Garde, S. Proc.Natl.Acad.Sci.USA {\bf 102}, 9475-9480  (2005).
\item $^{29}$D.Chandler,  Nature {\bf 437}, 640-7 (2005).
\item $^{30}$I.T.S.Li and G.C.Walker, G.C.  PNAS {\bf 108}, 16527-16532 (2011).
\item $^{31}$S.Garde and A.J.Patel, PNAS (2011) {\bf 108}, 16491-16492.
\item $^{32}$Y.S.Djikaev and E.Ruckenstein,  J. Phys. Chem. B  {\bf DOI: 10.1021/jp312631c} (2013).
\item $^{33}$F.M.Kuni, {\it Statistical Physics and Thermodynamics} (in Russian) (Nauka, Moscow, 1981).
\item $^{34}$R.Evans, in {\it Fundamentals
of inhomogeneous fluids}, ed. D. Henderson (Marcel Dekker, New York, 1992). 
\item $^{35}$D.E.Sullivan, Phys.Rev. B {\bf 20}, 3991-4000 (1979). 
\item $^{36}$P.Tarazona and R.Evans, Mol. Phys. {\bf 48}, 799-831 (1983).
\item $^{37}$P.Tarazona,  Phys.Rev.A  {\bf 31}, 2672-2679 (1985); {\it 32}, 3140 (1985) (erratum).
\item $^{38}$P.Tarazona, U.M.B.Marconi, and R.Evans,  Mol.Phys. {\bf 60}, 573-579 (1987).
\item $^{39}$Y.S.Djikaev and E.Ruckenstein,  J. Phys. Chem. B {\bf 116 }, 2820-2830 (2012).
\item $^{40}$E.Ruckenstein and Y.S.Djikaev, J. Phys. Chem. Lett. {\bf 2}, 1382-1386 (2011). 
\item $^{41}$Y.S.Djikaev and E.Ruckenstein,   
 Curr Opin Colloid Interface Sci. (2011) {\bf 16}, 272; doi:10.1016/j.cocis.2010.10.002 
\item $^{42}$N.F.Carnahan and K.E.Starling, J. Chem. Phys. {\bf 51}, 635-6 (1969).
\item $^{43}$J.D.Weeks, D.Chandler,  and H.C.Anderson, J. Chem. Phys. {\bf 54}, 5237-47 (1971).
\item $^{44}$R.Godawat, S.N.Jamadagni, and S.Garde, PNAS {\bf 106}, 15119-15124 (2009).
\item $^{45}$R.Fernandez-Prini and R.Crovetto, R.  J.Phys.Chem.Ref.Data  {\bf 18}, 1231 (1998).
\item $^{46}$D.Paschek, D.  J.Chem.Phys. (2004) {\bf 120}, 6674.
\item $^{47}$B.Guillot and Y.Guissani, J.Chem.Phys. {\bf 99}, 8075-8094 (1993).

\end{list}

\newpage 
\subsection*{Captions} 
to  Figures 1 to 5 of the manuscript 
{\sc
`` Temperature effect on the small-to-large crossover lengthscale of 
hydrophobic hydration "
}  by {\bf Y. S. Djikaev} and  {\bf E. Ruckenstein}.  
\subsubsection*{}
Figure 1. Density profiles near a spherical hydrophobe of  radius $R$ and $\eps_{\mbox{\tiny
wp}}/\eps_{\mbox{\tiny ww}}=0.75$. Different figure panels show results for different radii
($R/\eta=1,3,5,7,10,15,30,100$).  In each panel, different curves correspond to different temperatures
($T=293.15, 303.15, 313.15, 323.15, 333.15$ from bottom to top, respectively). The imposed chemical potential is
$\mu=-11.5989$ $k_BT_1$ (see the text for more details). 
\vspace{0.3cm}\\
Figure 2.  
Free energy of hydration and its entropic and
enthalpic constituents as functions of temperature, for hydrophobes of different radii with
$\eps_{\mbox{\tiny wp}}/\eps_{\mbox{\tiny ww}}=0.75$.
The curves correspond to  
$R/\eta=1,3,5,7,10,15,20,30,50$, and $100$ (from bottom to top in Figs.2a and 2c, from top to bottom in Fig.2b) 
The hydration free energy and its entropic and enthalpic components are shown as the dimensionless quantities 
$\overline{\De\Omega}, \overline{\Phi_S}$, and $\overline{\Phi_E}$, respectively (see the text for definitions). 
\vspace{0.3cm}\\
Figure 3.  
Free energy of hydrophobic hydration  $\De F_{\mbox{\tiny }}$  and its enthalpic and entropic constituents as 
functions of the hydrophobe radius $R$ with  $\eps_{\mbox{\tiny wp}}/\eps_{\mbox{\tiny ww}}=0.75$. The hydration
free energy and its constituents are shown as the dimensionless quantities  $\overline{\De\Omega},
\overline{\Phi_S}$, and $\overline{\Phi_E}$.  The curves correspond to five different temperatures,  $T=293.15,
303.15, 313.15, 323.15, 333.15$ (from bottom to top). 
\vspace{0.3cm}\\
Figure 4. Temperature dependence of the small-to-large srossover radii  $R_S^{0}$, $R_E^{0}$ and $R_{SE}$ 
for the hydration of solutes of given hydrophobicity $\eps_{\mbox{\tiny wp}}/\eps_{\mbox{\tiny ww}}=0.75$. 
\vspace{0.3cm}\\
Figure 5.   Comparison of the temperature dependence of the hydration free energy predicted by the combined
PHB/DFT model for small scale hydrophobes with both experimental data and simulational results for  simple
solutes. The three short solid lines are the model predictions  for solutes of  radii $R/\eta=1,\,3,\,5$ (from
bottom to top, respectively) with  $\eps_{\mbox{\tiny wp}}/\eps_{\mbox{\tiny ww}}=0.75$.  The filled circles and
squares represent the experimental data for xenon and methane, respectively.$^{45}$ The results by MD
simulations of neon, argon, krypton, xenon, and methane molecules in the SPCE model water are shown as
short-dashed, long-dashed, short-long-dashed, dash-dotted, and dotted curves, respectively.$^{46}$ The hydration
free energy and its entropic and enthalpic components are shown as the dimensionless quantities 
$\overline{\De\Omega}, \overline{\Phi_S}$, and $\overline{\Phi_E}$, respectively (see the text for details) 

\newpage
\begin{figure}[htp]
\begin{center}\vspace{0cm}
\includegraphics[width=13.5cm]{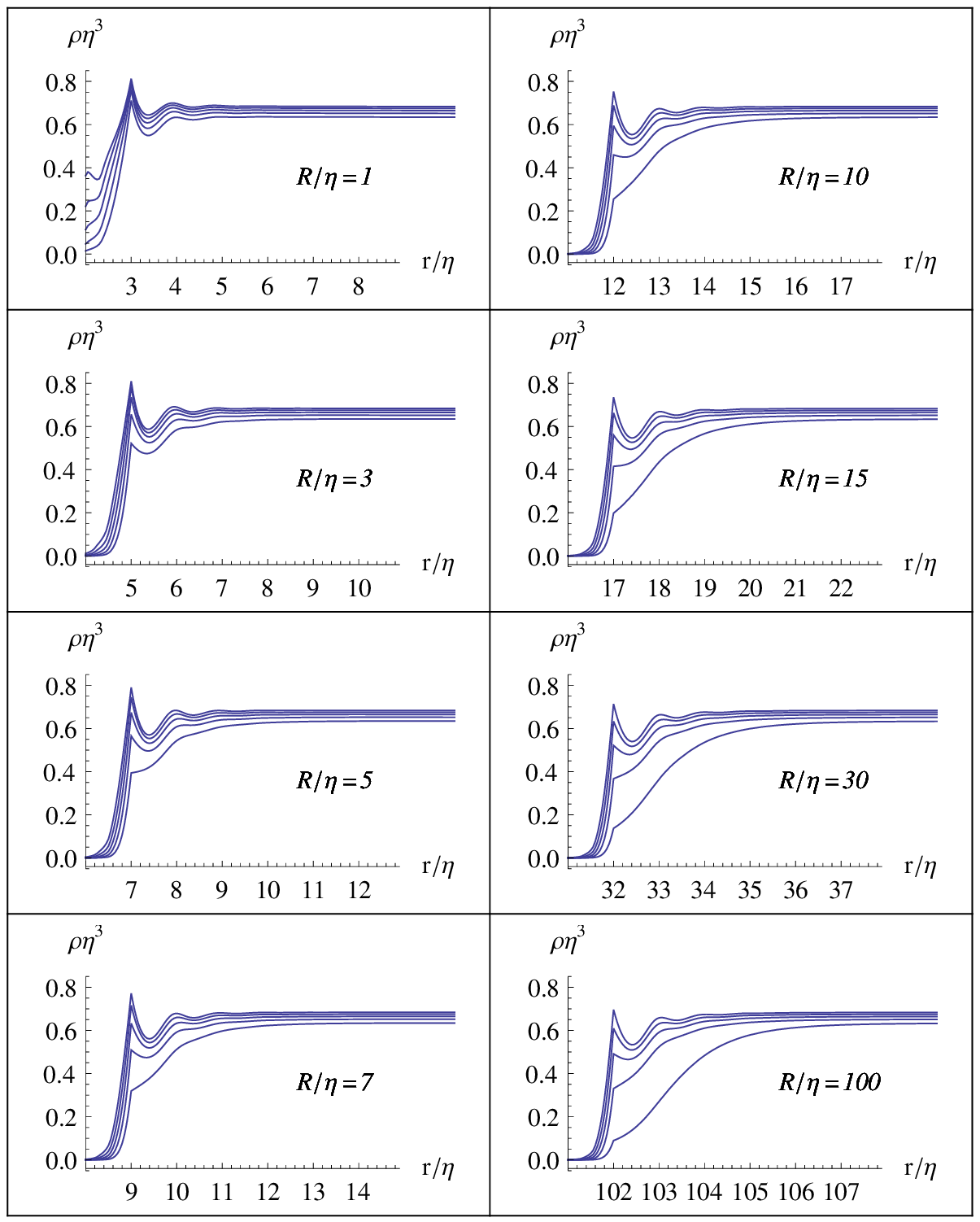}\\ [3.7cm]
\caption{\small }
\end{center}
\end{figure}

\newpage
\begin{figure}[htp]\vspace{-1cm}
	      \begin{center}
$$
\begin{array}{c@{\hspace{0.3cm}}c} 
              \leavevmode
      	      \vspace{-0.8cm}
	\leavevmode\hbox{a) \vspace{3cm}} &   
\includegraphics[width=8.7cm]{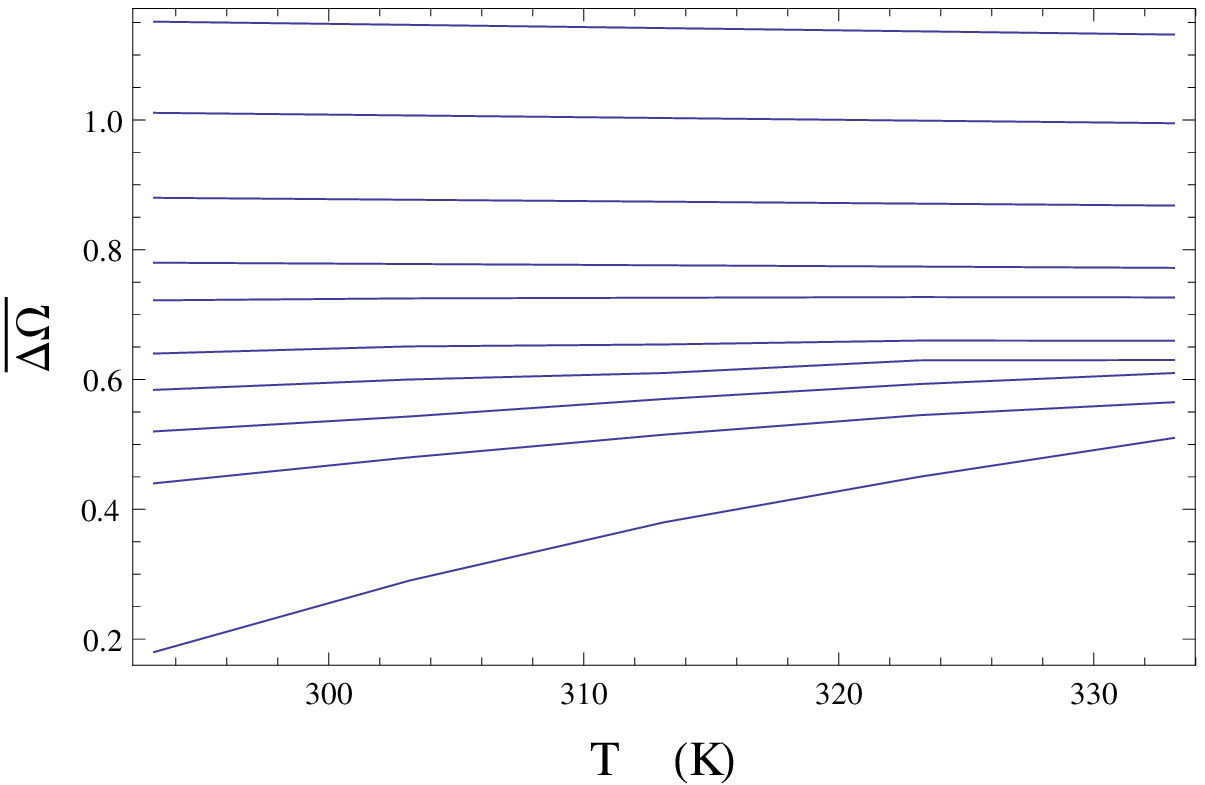}\\ [1.9cm] 
      	      \vspace{0cm}
	\leavevmode\hbox{b) \vspace{3cm}} &  
      	      \vspace{0.0cm}
\includegraphics[width=8.7cm]{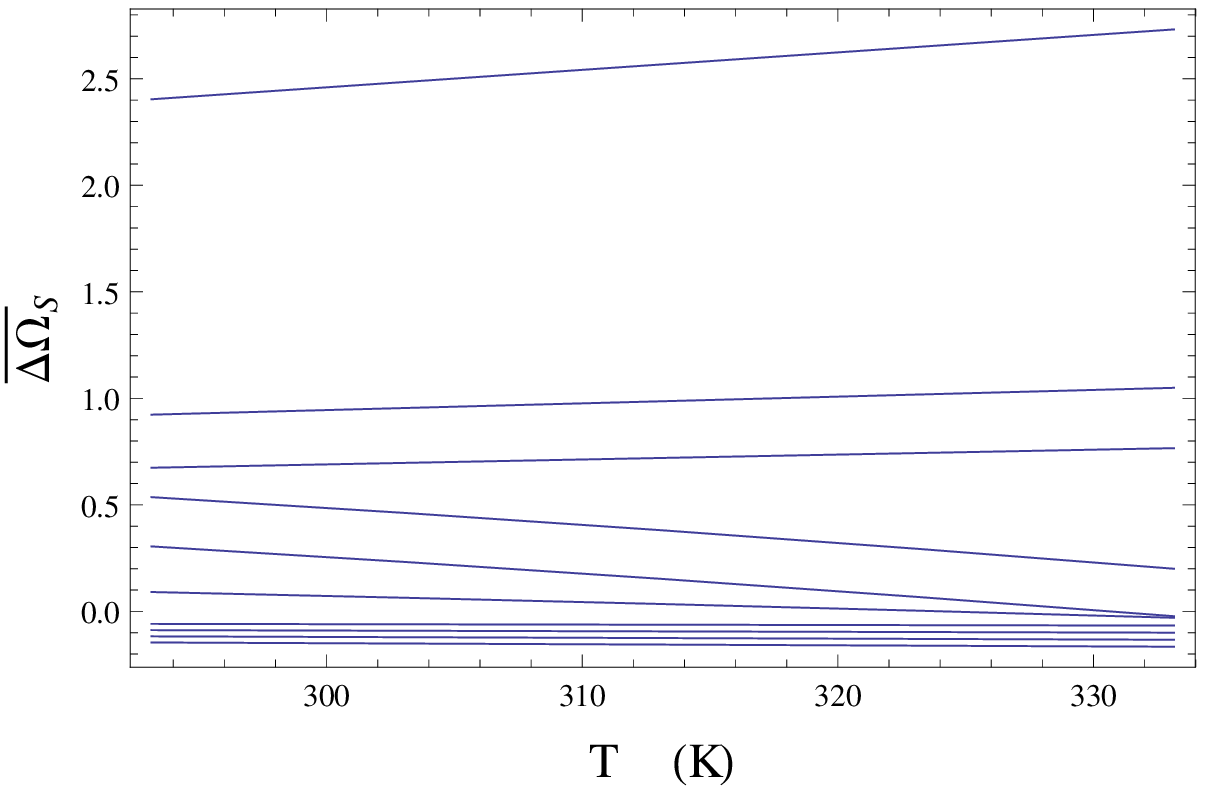}\\ [0.7cm] 
      	      \vspace{0cm}
	\leavevmode\hbox{c) \vspace{3cm}} &  
      	      \vspace{0.0cm}
\includegraphics[width=8.7cm]{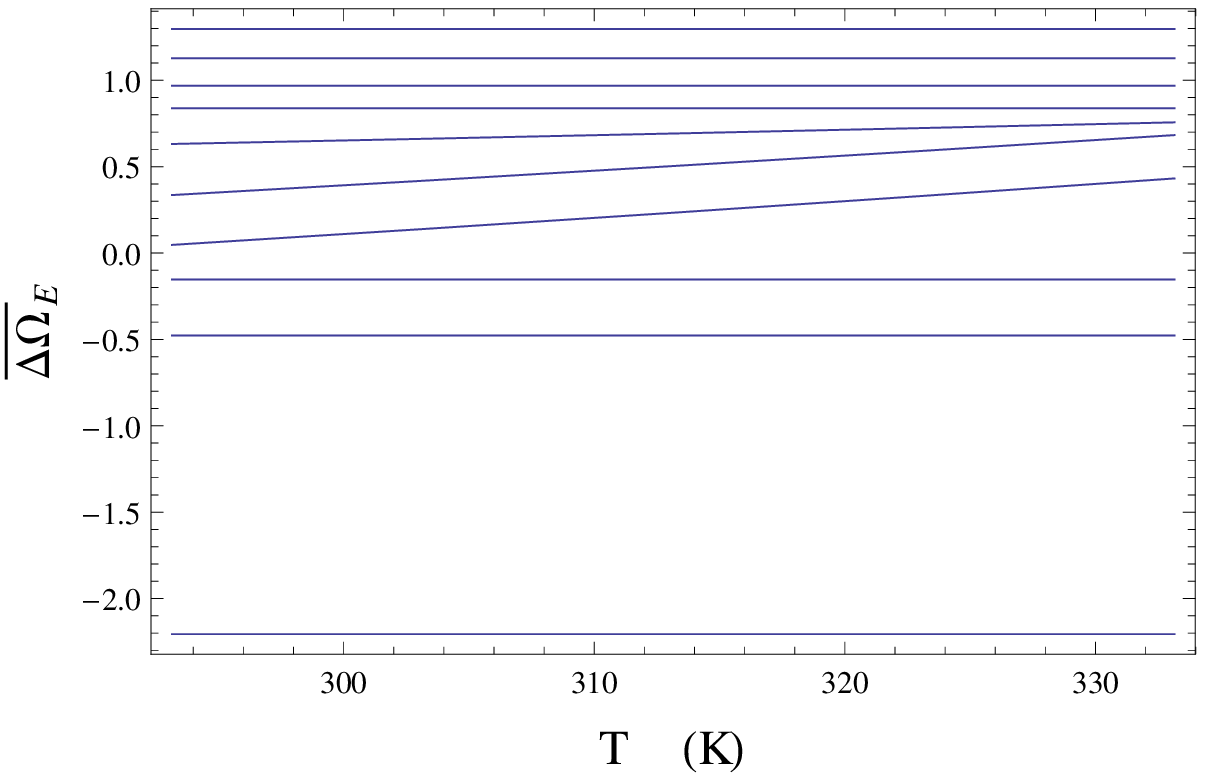}\\ [0.7cm] 
\end{array}  
$$  

	      \end{center} 
            \caption{\small } 
\end{figure}

\newpage
\begin{figure}[htp]\vspace{-1cm}
	      \begin{center}
$$
\begin{array}{c@{\hspace{0.3cm}}c} 
              \leavevmode
      	      \vspace{-0.8cm}
	\leavevmode\hbox{a) \vspace{3cm}} &   
\includegraphics[width=8.7cm]{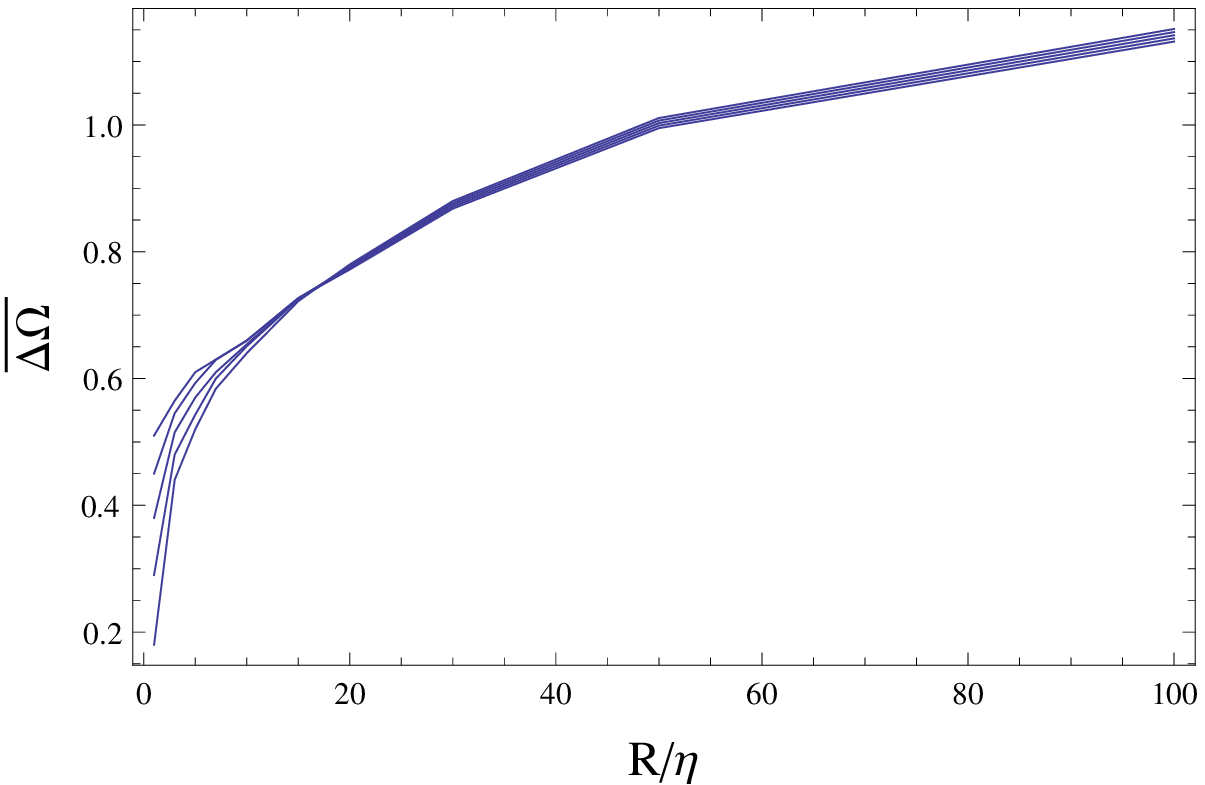}\\ [1.9cm] 
      	      \vspace{0cm}
	\leavevmode\hbox{b) \vspace{3cm}} &  
      	      \vspace{0.0cm}
\includegraphics[width=8.7cm]{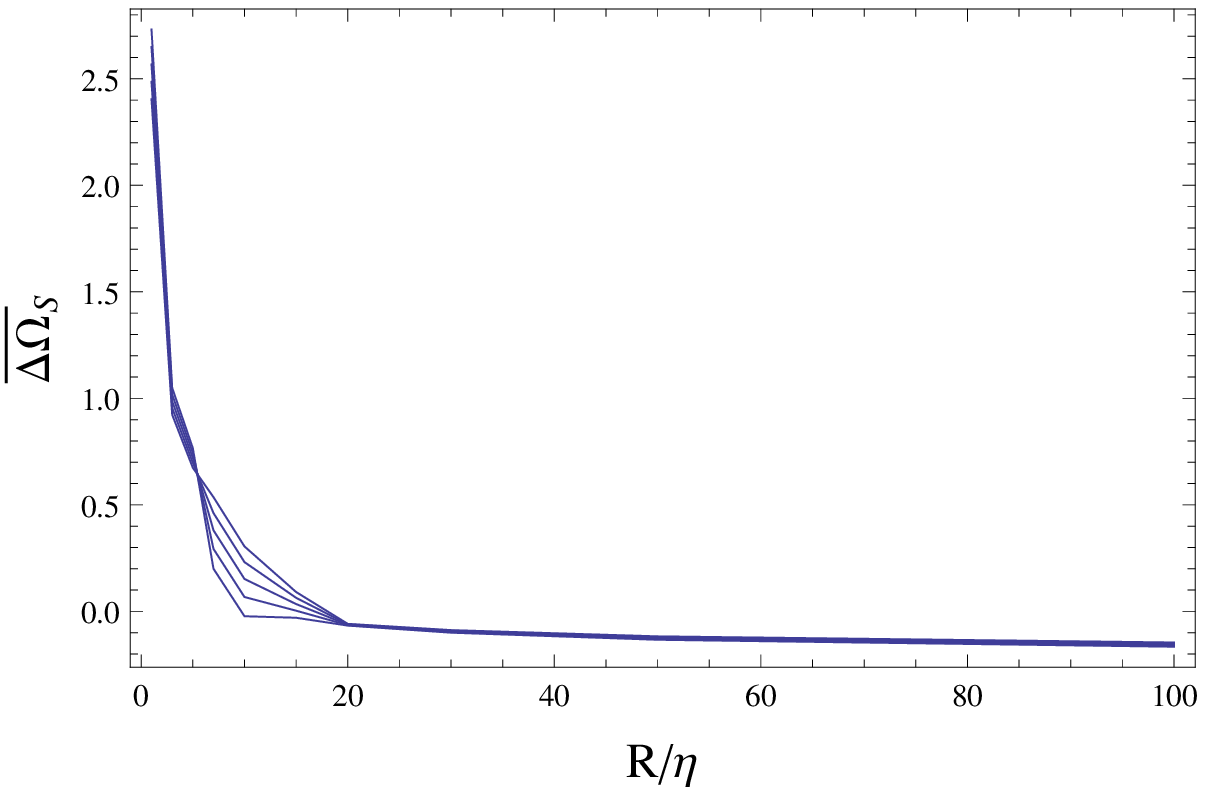}\\ [0.7cm] 
      	      \vspace{0cm}
	\leavevmode\hbox{c) \vspace{3cm}} &  
      	      \vspace{0.0cm}
\includegraphics[width=8.7cm]{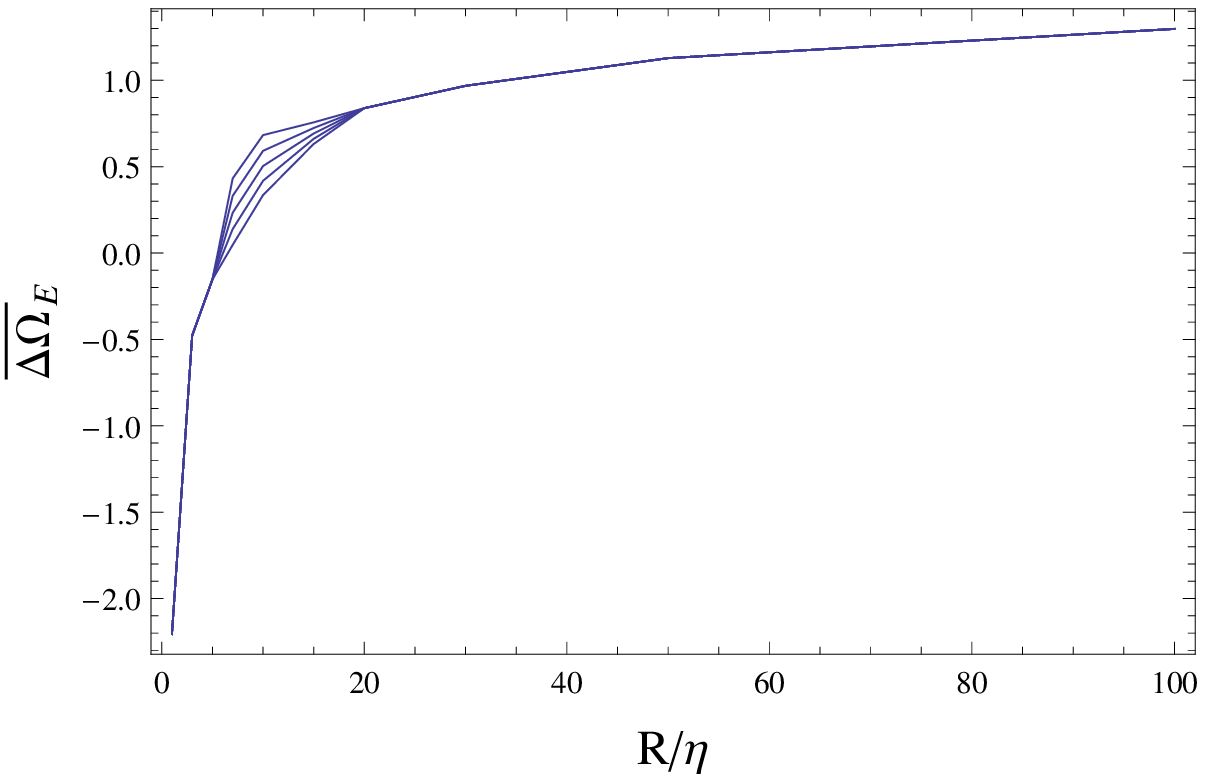}\\ [0.7cm] 
\end{array}  
$$  

	      \end{center} 
            \caption{\small } 
\end{figure}

\newpage
\begin{figure}[htp]
\begin{center}\vspace{1cm}
\includegraphics[width=8.9cm]{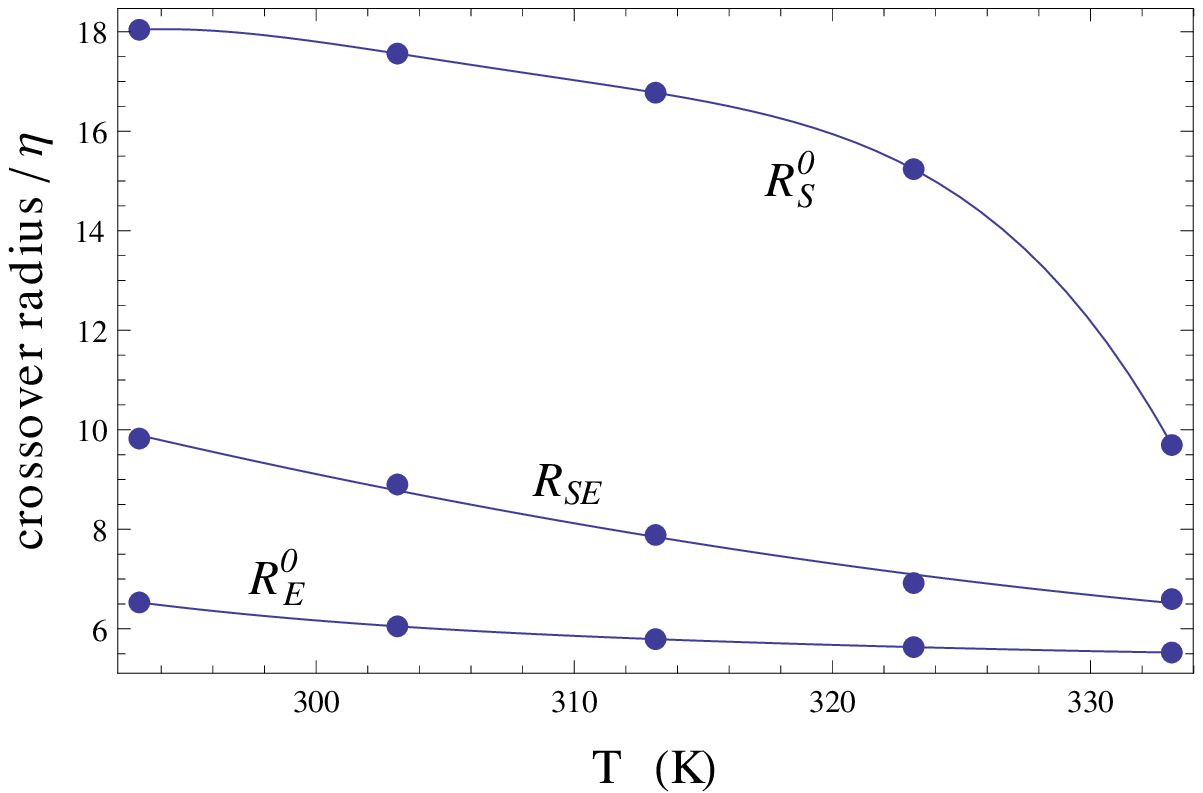}\\ [3.7cm]
\caption{\small }
\end{center}
\end{figure} 

\newpage
\begin{figure}[htp]
\begin{center}\vspace{1cm}
\includegraphics[width=8.9cm]{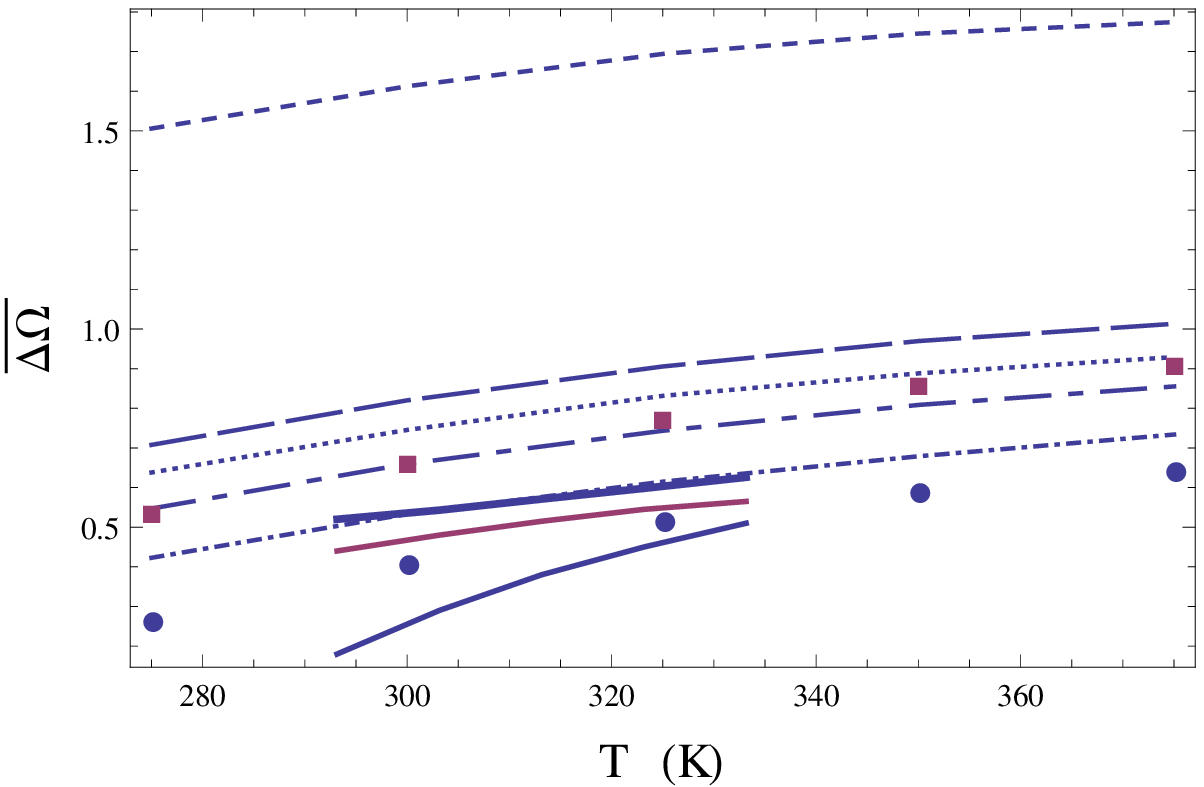}\\ [3.7cm]
\caption{\small }
\end{center}
\end{figure}

\end{document}